\documentclass[aps, reprint, prfluids, onecolumn, superscriptaddress, floatfix]{revtex4-2}

\usepackage{bm}% bold math
\usepackage{xcolor}
\usepackage{graphicx}
\usepackage{physics}
\usepackage{upgreek}
\usepackage[shortlabels]{enumitem}
\usepackage{siunitx} % dimensioned quantities
\DeclareSIUnit{\pixel}{px}
\DeclareSIUnit{\px}{px}
\DeclareSIUnit{\frame}{frame}
\DeclareSIUnit\litre{l} % redefine litre to have a small l
\usepackage{circledsteps}

\newlength{\figwidth}
\newcommand{\crit}{{\alpha}}
\newcommand{\mach}{{M}}
\newcommand{\cphi}{{c}}

\newcommand{\vcap}{{\ensuremath{v_\text{c}}}} % capillary speed v_c or v_cap or v_gamma
\newcommand{\hlim}{{\ensuremath{h_\infty}}}

\newcommand{\area}{{\ensuremath{\upsigma}}} % ou a ou s ou \Sigma ou \sigma ou \upsigma
\newcommand{\mucl}{{\ensuremath{\mu_\text{cl}}}}
\newcommand{\mutot}{{\ensuremath{\mu_\text{tot}}}}

\newcommand{\muadim}{{\ensuremath{\overline{\mu}}}}
\newcommand{\mucladim}{{\ensuremath{\overline{\mucl}}}}
\newcommand{\mutotadim}{{\ensuremath{\overline{\mutot}}}}
\newcommand{\Badim}{{\ensuremath{\overline{B}}}}

\newcommand{\ut}{{\ensuremath{u_t}}}
\newcommand{\un}{{\ensuremath{u_n}}}

\newcommand{\leftsemicirc}{{\put(3.5,2.5){\oval(4,4)[l]}\phantom{\circ}}}
\newcommand{\rightsemicirc}{{\put(1.5,2.5){\oval(4,4)[r]}\phantom{\circ}}}

\newcommand{\ReN}{\ensuremath{\mathrm{Re}}} % capillary number
\newcommand{\lcap}{{\ensuremath{\ell_\text{c}}}} % capillary speed v_c or v_cap or v_gamma
\newcommand{\stress}{\tau}

\newcommand{\vunit}[1]{{\ensuremath{\vb{\hat{#1}}}}} % unit vector
\newcommand\bnabla{\boldsymbol{\nabla}}
\newcommand\bcdot{\boldsymbol{\cdot}}

\makeatletter
\newcommand*{\defeq}{\mathrel{\rlap{%
    \raisebox{0.3ex}{$\m@th\cdot$}}%
\raisebox{-0.3ex}{$\m@th\cdot$}}%
    =}
\newcommand*{\eqdef}{=\mathrel{\rlap{%
    \raisebox{0.3ex}{$\m@th\cdot$}}%
\raisebox{-0.3ex}{$\m@th\cdot$}}%
}
\makeatother

\newcommand{\eps}{{\ensuremath{\upepsilon}}} % small parameter
\usepackage{todonotes} % add to do notes with \todo{foo}
\usepackage{soul}
\sethlcolor{orange}

\newcommand{\todohl}[2]{%
    \makeatletter\if@todonotes@disabled%
    #1%
    \else%
    \hl{#1}\todo{#2}%
    \fi\makeatother%
}

\colorlet{revision1color}{black}
\colorlet{revision2color}{black}

\newcommand{\revisecross}[1]{}

\begin{document}

\title{%
    Viscous Bending Mitigates the Spontaneous Meandering of Rivulets in Hele-Shaw Cells%
}

\author{Grégoire Le Lay}
\email{gregoirelelay@protonmail.com}
\affiliation{%
    Matière et Systèmes Complexes UMR\,7057, Université Paris Cité, CNRS,\\
    F-75013 Paris, France%
}%
\affiliation{%
    Laboratory of Fluid Mechanics and Instabilities, EPFL,\\
    CH-1015 Lausanne, Switzerland
}%
\author{Adrian Daerr}
\affiliation{%
    Matière et Systèmes Complexes UMR\,7057, Université Paris Cité, CNRS,\\
    F-75013 Paris, France%
}%

\date{\today}

\begin{abstract}
    We investigate the spontaneous meandering of slender rivulets in Hele–Shaw cells and identify the physical mechanism that selects the most unstable wavenumber,
    a quantity that has remained elusive even since the identification of the instability threshold [Daerr \textit{et al.}, \textit{Phys. Rev. Lett.} 106, 184501 (2011)].
    Earlier criteria did not distinguish between wavelengths and thus predicted an undiscriminated amplification of arbitrarily short perturbations.
    By incorporating viscous bending into the depth-averaged Navier-Stokes equations,
    we show that this effect is responsible for the selection of a fastest-growing mode,
    answering a question that has remained open for 15 years.
    We answer the open question of whether the meandering instability is absolute or convective.
    Our analysis also provides a simpler alternative derivation of the instability criterion,
    based on a low-viscosity assumption, and finally it yields a new physical interpretation of the mechanism:
    the destabilization arises directly from friction effects, instead of being caused by inertial forces.
    Together, these results complete the linear-stability picture of rivulet meandering in confined geometries,
    and establish viscous bending as a key parameter governing wavelength selection.
    They lay the groundwork for future exploration of the nonlinear features of the spontaneous meandering instability.
\end{abstract}

\maketitle

\section{Introduction}\label{sec:intro}

When a liquid flows down an inclined plane,
it tends to organize into slender, continuous streams termed rivulets,
which, above a certain flow rate, weave back and forth, adopting a sinuous, oscillating trajectory.
This meandering instability is governed by the interplay between gravity, inertia, surface tension,
and the wetting properties of the underlying substrate.
Understanding the mechanics of these meandering flows is of course of fundamental interest,
but it is also useful in several industrial processes:
it is necessary in order to optimize mass and heat transfer in industrial heat exchangers,
design uniform liquid coating processes,
and manage water runoff on architectural structures or vehicle windshields.

Using a liquid that perfectly wets the solid substrate,
and flowing it between two parallel plates (a Hele–Shaw cell),
generates meanders that can move upwards or downward.
This was first observed in the case where the fluid is a surfactant solution in water~\citep{anand1986, drenckhan2004, legrand-piteira2006a}.
Such an oscillating rivulet of surfactant solution in a Hele–Shaw cell is a potentially useful system for the macroscopic measurement
of the microscopic properties of these mixtures, which are usually difficult to measure, such as surface elasticity or
surface shear viscosity~\citep{drenckhan2007}.
However, the richness and complexity of the observed behaviors have motivated the search for a simpler system that could be better understood.

By replacing the water-surfactant mix with low-viscosity oils,
Daerr et al. (2011)~\cite{daerr2011} determined the threshold beyond which the sinuosity of the trajectory develops,
and showed that the general destabilization mechanism was valid for other configurations.
However, the model presented in this study does not make it possible to understand the typical wavelength of the oscillations:
according to this model, beyond the threshold all wavelengths become unstable at the same time,
including infinitely small wavelengths, which is not physical;
and the dispersion relation does not possess a maximum that would explain the selection of a particular mode.
The 2011 article calls for considering a nonlinear mechanism of wavelength selection,
but before embarking on such a study,
it is first necessary to derive a linear system that does not predict the indiscriminate amplification of all frequencies.

In the present paper, we derive a robust model for meandering rivulets,
which contains all necessary ingredients to solve the aforementioned problems,
and we use it in order to answer several oen questions about this system.
In section~\ref{sec:dyna}, we provide a rigorous derivation of the general equations for meandering rivulets
starting from depth-averaged Navier-Stokes equations,
detailing all the forces that act on the system,
listing all the necessary hypothesis,
and incorporating a previously overlooked ingredient, viscous bending.
In section~\ref{sec:linear} we explore a linear approximation of this model,
deriving the wavenumber-dependent instability criterion.
Notably, we prove that the instability is convective in the laboratory frame of reference,
and we provide a useful approximation for the growth rate of the instability.
In section~\ref{sec:physics}, we discuss the physical interpretation of the instability,
providing and detailing a new interpretation of the spontaneous amplification of meanders.
Last, in section~\ref{sec:multiscale},
we use a multiple scales approach in order to retrieve the previously obtained approximation for the growth rate,
in a way that can be generalized and adapted for non-linear studies.

A characteristic of the spontaneous meandering regime is that the rivulet always retains a constant width:
in the absence of acoustic forcing, width modulations are no longer coupled to the sinuous oscillations,
and in the remainder of this part we will therefore always consider the width to be constant: $w_0$.
The area of the transverse cross-section of the rivulet $\area_0$ is therefore also constant,
which implies in particular that
the mean velocity of the fluid in the rivulet $u_0$, as well as the phase velocity of the sinusoidal perturbations $\vcap$,
are also fixed:
we therefore assume that they remain constant throughout the dynamics.
In the models we use, these velocities thus depend only on the experimental conditions, in particular on the flow rate $Q$,
but never on dynamical variables (e.g.~the wavelength or the oscillation amplitude).

\section{Dynamical equations}\label{sec:dyna}

Let us first carefully establish the dynamical equations that encode the behavior of the rivulet,
precising at each step the hypothesis we make.

The fluid is injected between two vertical plates (which form an air-filled Hele-Shaw cell),
and falls down under the influence of gravity.
It forms a long \textsl{rivulet}, that is delimited on the side by two menisci.
Since the spacing $b$ between the front and back plates is smaller than the capillary length $\lcap \defeq \sqrt{\gamma/(\rho\, g)}$,
these menisci are (in first approximation) semi-circular (see fig. REF (center)).
The internal \textsl{width} of the rivulet is noted $w$,
and the cross-sectional area of the rivulet is thus $\area \defeq w\, b + (1-\pi/4)\,b^2$.
The horizontal position of the centerline of the rivulet (its \textsl{path} or \textsl{fish bone}) is noted $z(x)$, with $x$ being the vertical coordinate (see fig. REF (center)).
The rivulet is supposed only slightly inclined ($\partial_x z \ll 1$),
which guarantees that $z(x)$ is mono-valuated.

We consider the rivulet \textsl{slender} in the $(x,z)$ plane,
meaning that the radius of curvature of its path $1/\partial_{xx} z$ is always greater than the cell spacing $b$ and the rivulet width $w$.
The rivulet can thus be thought as a one-dimensional object,
and the goal of this section is to establish the equations that dictates its time evolution.
First, we need to perform a \textsl{gap averaging} of the full Navier-Stokes equations,
in order to describe the fluid dynamics that dominate in the bulk.
Then, we need to add several terms in order to take into account the effects of the rivulet geometry on the fluid velocity.

\begin{figure}[h]
    \centerline{\includegraphics[scale=1]{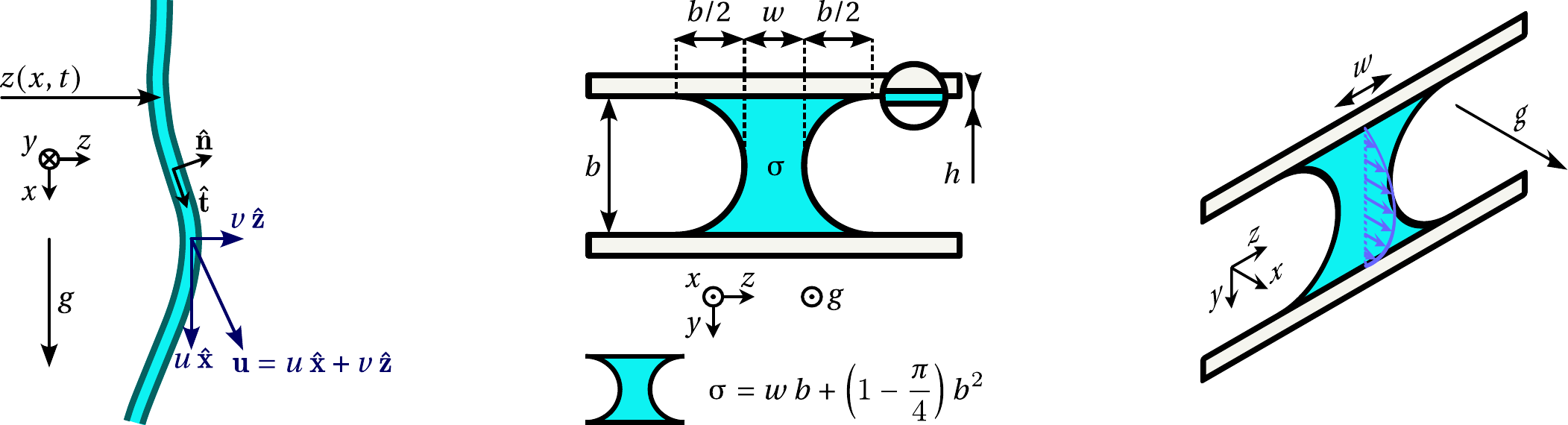}}
    \caption{
    (left) Notations used in this paper for the rivuklet position and the fluid velocity.
    (center) Geometry of the rivulet cross-section.
    (right) Parabolic flow profile.
    }
    \label{fig:drawing_notations}
\end{figure}

\subsection{Gap-averaged bulk description}

The fluid is confined in a Hele-Shaw cell,
meaning that the spacing between the plates is sufficiently small to keep the Reynolds number $\ReN = b\,U / \nu$
(with $U$ the typical flow speed) at moderate values.
We do not provide a precise criterion for what we consider a \textsl{moderate} Reynolds number,
except that it must be small enough so that to understand the behavior of the rivulet,
there is no need to account for any secondary flow
(such as pairs of counter-rotating vortices in the cross section). % Name drop the term `Dean vortices' ?
Using a lubrication approximation,
the flow in the bulk can then be entirely described using the gap-averaged fluid velocity $\vb{v}(x,z) \defeq \ev{\vb{v}_{3D}(x,y,z)}_y$.
At very low Reynolds numbers, the gap-averaged velocity $\vb{v}$ obeys the simple Darcy's law.

In order to account for inertia in our model,
we add the term $(\alpha\,\partial_t + \beta\,\vb{v}\bcdot \bnabla) \vb{v}$; leading to the following Navier-Stokes equation
being verified by the gap-averaged speed inside the bulk:
\begin{align}\label{eq:bulk}
    \rho\,(\alpha \, \partial_t + \beta \, \vb{v}\bcdot \bnabla) \vb{v}
    &= \rho\,\vb{g} - \frac{12\,\nu\,\rho}{b^2}\,\vb{v}\ .
\end{align}
We can use different values for the coefficients $\alpha$ and $\beta$, depending on the assumptions made.
A simple averaging of the Navier-Stokes equations across the gap (implicitly assuming a plug-flow) leads to $\alpha=\beta=1$.
Assuming a more realistic parabolic Poiseuille profile leads to $\alpha=1$ and $\beta=6/5$, as found by Gondret and
Rabaud~\cite{gondret1997}.
Later, Ruyer-Quil~\cite{ruyer-quil2001} showed that using any king of weighted residual methods leads to $\alpha=6/5$ and $\beta=54/35$.
In this paper we do not aim at obtaining precise numerical results, but to provide a deep qualitative understanding of the physical phenomenon at play:
we henceforth assume $\alpha=\beta=1$ in order to increase the readability of the following equations,
keeping in mind that any future quantitative comparison with experimental data should be done using more mathematically appropriate coefficients.

In order to link the local fluid velocity in the bulk to the displacement speed of the rivulet,
we perform a integration along the direction normal to the rivulet path (the $z$ direction for a rivulet falling straight down).
This leads to the equation
\begin{align}\label{eq:bulk-int}
\rho\,\area\,(\partial_t + \vb{u}\bcdot \bnabla) \vb{u}
&= \rho\,\vb{g}\,\area - \rho\,\area\,\mu \, \vb{u} + \vb{f}_{\text{geo}} \qqtext{where} \vb{u}(x,t) = \frac1\area\int_{(\area)} \vb{v}\, \dd{\area}
\end{align}
is the local speed of material points of the rivulet, as shown in figure~\ref{fig:drawing_notations}~(left),
$\mu$ is a coefficient which depends on the rivulet internal width ,
and $\vb{f}_\text{geo} = \vb{f}_c + \vb{f}_f + \vb{f}_b$ is the sum of supplementary forces dues to the rivulet geometry.

If the rivulet cross-section was rectangular and the flow profile was parabolic with $y$,
as shown in figure~\ref{fig:drawing_notations}~(right),
then we would have $\mu = \mu_\infty \defeq 12\,\nu/b^2$.
However, because of the semi-circular geometry of the air-liquid interfaces,
the velocity profile in the $(y,z)$ plane is more complicated and the Darcy coefficient $\mu$
changes with the rivulet width $w$~\cite{drenckhan2007}.
The values of $\mu(w) / \mu_{\infty}$ can be computed without much effort using, for example,
finite element techniques to solve the Laplace equation on the rivulet cross-section~\cite{drenckhan2007, lelay2025thesis}.
Qualitatively, as the flow rate augments, the rivulet grows wider,
and $\mu$ tends toward $\mu_\infty$ as the contribution of the flow in the menisci become negligible.

Concerning the velocity $\vb{u}$, note that there are two interesting bases on which it can be projected.
\begin{enumerate}
    \item The natural base (Frenet base), linked to the rivulet, in which $\vb{u} = \ut \, \vunit{t} + \un \, \vunit{n}$,
    where $\vunit{t}$ is a vector tangential to the rivulet path,
    and $\vunit{n}$ is normal to the said path and oriented toward the right.
    Note that these base vectors change with both space and time, which makes them highly unpractical to use.
    \item The static base (linked to the plate, fixed in the laboratory frame of reference),
    in which $\vb{u} = u \, \vunit{x} + v \, \vunit{z}$,
    where $\vunit{x}$ points down (in the direction of gravity)
    and $\vunit{z}$ points right (in the transverse direction, parallel to the plate).
\end{enumerate}
The advective derivative is for example can be written $ \vb{u}\bcdot \bnabla = \ut \, \partial_s $,
with $\partial_s = \cos\theta \, \partial_x$ being the derivative along the curvilinear coordinate.

\subsection{Supplementary geometric forces}

Equation~\eqref{eq:bulk-int} is incomplete because it only encodes the effect of the bulk of the fluid on the dynamics.
In order to understand the behavior of the rivulet,
we need to take into account supplementary effects that are generated by the geometry of a rivulet with constant width.
The object of this subsection is to describe the three main effects: capillary restoring force,
contact-line friction, and viscous bending force.

\subsubsection{Capillary effects}

Since the rivulet is bounded by menisci,
the geometry of which being mainly driven by air-liquid surface tension,
capillarity plays a dominant role in its dynamics.
The most direct way to understand the effects of surface tension is to consider the change of capillary pressure
inside the rivulet that is caused by the curvature of the menisci.

Let us first consider individually one of the menisci bounding the rivulet.
At lowest order, since the cell spacing is smaller than the liquid’s capillary length,
such a meniscus exhibits a semicircular geometry in the $(y,z)$ plane, with curvature $2/b$.
This implies a change of pressure $ \Delta P = -{2\,\gamma}/{b}$ in the liquid with respect to the surrounding fluid.

\begin{figure}[h]
    \centerline{\includegraphics[scale=1]{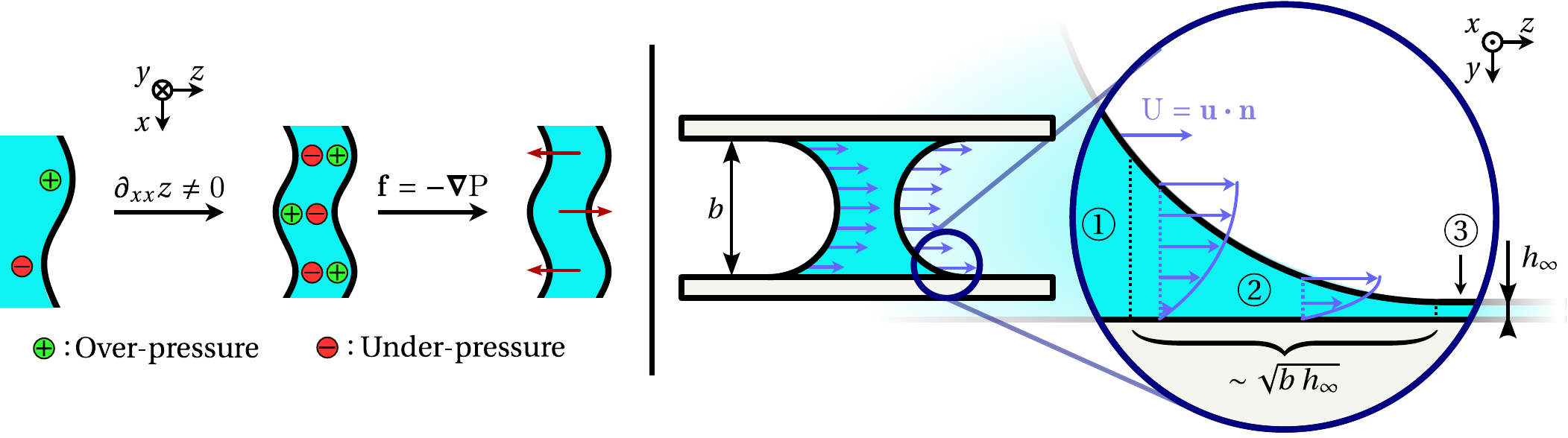}}
    \caption{
        (left) MEchanism giving birth to the transverse capillary force.
    \\
    (right) Origin of the viscosity associated with transverse motions.
    In the reference frame of the plates, the interface moves in the $z$ direction at a finite velocity.
        The no-slip condition at the plates implies a velocity gradient that diverges as the fluid thickness tends to \num{0}.
        This divergence is regularized by the existence of a finite prewetting film of thickness $h_\infty$.
    }
    \label{fig:drawing_effects}
\end{figure}

In reality, the meniscus interface in the $(y,z)$ plane is not exactly a semicircle.
The meniscus is slightly deformed because the trace of the liquid--air interface in the cell plane $(x,z)$ is curved.
This changes the value of the capillary pressure jump.
The multiplicative factor associated with this correction was computed by Park \& Homsy~\cite{park1984} using matched asymptotic expansions,
in the limit of small velocities.
Thus, considering a liquid--air interface that describes a curve $\zeta(x)$ in the $(x, z)$ plane,
the pressure jump inside the liquid is
\begin{align}
    \Delta P = -\frac{2\,\gamma}{b}\qty(1 - \frac{\pi}{4} \frac{b}{2} \, \partial_{xx}\zeta)
\end{align}
which means that the pressure inside the rivulet is a function of the curvature, as shown on figure~\ref{fig:drawing_effects}~(left).

Let us write the difference between the capillary pressure jumps due to the two interfaces on each side of the rivulet (to first order in $z$):
\begin{align}
    \Delta P_\text{int} = \frac{\pi\,\gamma}{4}\big(\partial_{xx}z_\rightsemicirc - \partial_{xx}(-z_\leftsemicirc)\big)
    = -\frac{\pi\,\gamma}{2}\,\partial_{xx} z
\end{align}
where $z_\rightsemicirc = z - w/2$ and $z_\leftsemicirc = z + w/2$ represent the positions of the left and right menisci, respectively.
Note the change of sign of the curvature terms, which accounts for the inversion of the liquid / air phases.
These capillary pressures are shown on figure~\ref{fig:drawing_effects}.
This means that on a small portion of rivulet of length $\delta\ell$,
the rivulet experiences a force due to Laplace pressure directed along the normal to its trajectory,
with magnitude $b\,\delta\ell \, \Delta P_\text{int}$.
The corresponding force per unit length, divided by the fluid density, is then
\begin{align}
    \vb{f}_c = \frac{\pi\,\gamma\, b}{2}\,\partial_{xx} z
    = \rho\,\area\, \vcap^2\,\partial_{xx}z
    \qqtext{with} \vcap = \sqrt{\frac{\pi\,\gamma\, b}{2\,\rho\, \area}}\quad\text{the \textsl{capillary speed}.}
\end{align}.

This force is straightforward to interpret:
it corresponds to a capillary restoring force that opposes the increase of the menisci surface area
caused by the curvature of the rivulet’s path.

Since our model now contains a capillary restoring force as well as inertia,
we naturally expect \textsl{transverse capillary waves} to propagate on the rivulet.
As we will see later, such transverse waves travel at the \textsl{capillary speed} $\vcap$
in the reference frame of the rivulet~\cite{lelay2026}.

Note that in this system can also sustain the propagation of
\textsl{longitudinal} capillary waves~\cite{lelay2025, lelay2026}.
However, since no mechanism here can inject energy into the longitudinal modes of deformation,
and thus the rivulet always conserve the same width through spontaneous meandering,
we choose to omit here for the sake of simplicity and concision.

\subsubsection{Contact-line supplementary friction}

When the rivulet moves in the direction perpendicular to its path,
the meniscus interface in the transverse plane keeps the same geometry
(at least in first approximation).
This means that each point of the free surface of the rivulet moves relative to the plates with a finite velocity $\un = \vb{u}\bcdot \vunit{n}$.
However, by assumption the fluid velocity is zero at the walls~:
this implies the existence of a velocity gradient in the $y$ direction between the plate and the interface.
As the meniscus becomes thinner, the shear increases
(see figure~\ref{fig:drawing_effects}~(right)),
eventually diverging in the limit where the menisci touch the plates.
It is therefore important to take into account the fact that,
due to total wetting between the liquid and the plates,
the rivulet always \textsl{slides} on a liquid film of thickness $\hlim$.

To incorporate into our model this strong increase in dissipation in the thin, moving parts of the menisci,
we must add a corrective term to our equation, represented by the friction force $\vb{f}_f$.
Near the menisci, the velocity is essentially normal to the contact line~\cite{rio2005}.
This corrective term must therefore depend on the velocity normal to the path $\un$.
To estimate it at the level of orders of magnitude, we use the following reasoning:
when integrating the viscous stress term of the Navier--Stokes equation in the $y$ direction,
three regions can be distinguished (see figure~\ref{fig:drawing_effects}~(right))~:
\begin{itemize}
    \item Region $\Circled{1}$, centered on the position of the rivulet $z(x,t)$ and of width $w/2$,
    where the liquid spans the entire cell gap and the flow is of Poiseuille type;
    \item region $\Circled{2}$, comprising the menisci, where dissipation takes place;
    \item finally region $\Circled{3}$, outside the rivulet, consisting of a microscopic film of thickness $\hlim$.
\end{itemize}

In region $\Circled{3}$, the viscous stress inside the nearly motionless film
is very small compared to that occurring in the rivulet.
This contribution can be neglected.

In region $\Circled{1}$,
the flow is to first approximation a Poiseuille flow for a cell of thickness $b$,
with an average velocity along $z$ denoted $\un$.
The viscous stress at the plates is then
$\stress_{yz} = -\rho \nu \frac{6\un}{b}$.
The total force per unit length of the rivulet exerted by the plates on the fluid is
$2\,w\, \stress_{yz} = -\rho\,b\,w\,\mu_\infty\,\un$ with $\mu_\infty = 12 \nu / b^2$~:
the viscous dissipation is described by the Darcy law discussed previously.

In region $\Circled{2}$,
the viscous stress at the plates is
$\stress_{yz} = -\rho \nu \frac{6\un}{h(z)}$,
where $h(z)$ is the meniscus thickness at the considered abscissa $z$.
Since this stress diverges as $1/h(z)$,
the total dissipation is dominated by that occurring in the thinnest parts of the meniscus,
where $h(z)$ is of the order of $\hlim$ and therefore
$\stress_{yz} \approx -\rho \nu \frac{6\un}{\hlim}$.
Roughly estimated,
the typical size of this region is of the order of $\sqrt{b\,\hlim}$
(this is the distance over which the meniscus thickness,
assimilated to a semicircle of radius $b/2$,
increases from $\hlim$ to $2\,\hlim$).

Thus, the total force (per unit length of the rivulet) exerted by the plates on the fluid
at the four locations where the menisci are thinnest is
$\vb{f}_f\bcdot\vunit{n} = 4\, \sqrt{b\,\hlim}\, \stress_{yz} = -\rho\,\area\,\mucl\,\un$
with $\mucl \sim 2\,\mu_\infty\,\frac{b^2}{\area}\,\sqrt{\frac{b}{\hlim}}$.
A more rigorous calculation involving the integration of the stress while varying the height $h(z)$
makes it possible to obtain a more precise estimate of the prefactor,
which remains of order unity~\cite{cantat2013, bazarmucl}.

The difficult point is the evaluation of the film thickness $\hlim$,
which intervenes directly in the definition of $\mucl$.
Theoretically, this thickness is not the same depending on whether one considers
the film left behind the rivulet,
which depends on the velocity of the receding meniscus,
or the film already formed on which the rivulet advances.
The latter results from previous wetting events,
and strictly speaking its thickness therefore depends locally on the history of the rivulet motion.
In this article, we will always make the simplifying assumption
that the rivulet slides on a film whose thickness is everywhere constant,
which leads us to neglect spatial and temporal variations of $\mucl$.
This effective thickness $h_\infty$ is taken of the order of a few micrometers,
in accordance with recent experimental measurements~\cite{lelay2025a}.

\subsubsection{Viscous bending}

Up until now,
all the physical ingredients we introduced in order to build our model have already been used
to describe the behavior of spontaneously meandering fluid rivulets.
Together, they allow one to compute the threshold of the meandering instability,
but they fail at explaining why all wavelength are not attenuated~\cite{daerr2011}.
In order to introduce a small wavelength cut-off,
we need to take into account a new physical effect: viscous bending of the rivulet.

\begin{figure}[h]
    \centerline{\includegraphics[scale=1]{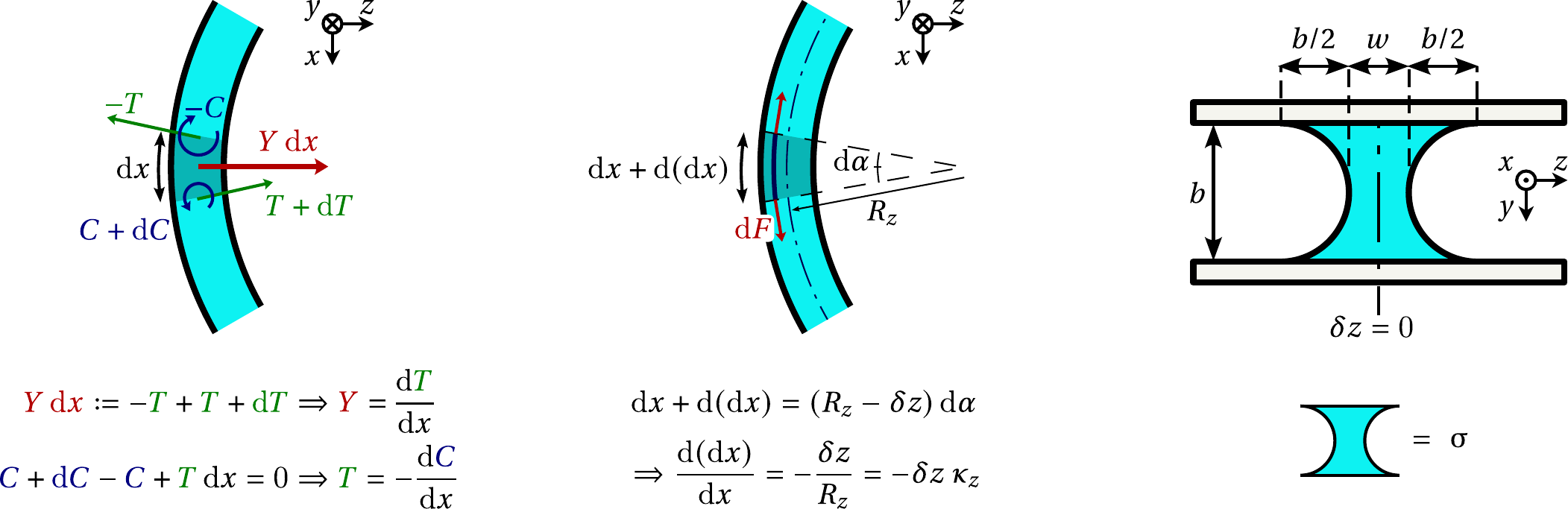}}
    \caption{
        (left) Relation between the internal bending moment $C$, the shear force $T$ and the lineic bending force $Y=\vb{f}_b\bcdot\vunit{n}$.
        (center) The neutral axis of the beam (discntinuous line) keeps the same length,
        while non-neutral axes are deformed (here in dark blue, the axis is lengthened).
        (right) cross-section of the rivulet,
        with the notations used to compute the planar quadratic moment of area $I_{y}$ in annex~\ref{sec:quadmom}.
    }
    \label{fig:drawing_bending}
\end{figure}

Since the rivulet is a slender object subject to small deformations which do not modify its cross-sectional area $\area$,
it is analogous to a \textsl{liquid beam} in the sens of continuum mechanics.
When this homogeneous beam is bent, it is subject to a transverse restoring force that mitigates the deformation.
In the case of solid beams, this force is of elastic origin.
For liquid beams, it is a viscous force, that opposes the change of curvature rather than the curvature itself~\cite{buckmaster1973, buckmaster1975, ribe2006, lemerrer2012}.
This bending force $\vb{f}_b\bcdot\vunit{n}\,\dd{x}$ is related to the internal bending moment $C$ by the relation
\begin{align}
    \vb{f}_b\bcdot\vunit{n} = -\pdv[2]{C}{x}
\end{align}
as can be seen in figure~\ref{fig:drawing_bending}~(left).
We can express the bending moment in the form
\begin{align}
    C = -\iint_{(\area)} \delta z\dd{F}
\end{align}
where $\area$ is the cross-sectional area of the rivulet (in the $(y,z)$ plane),
$\delta z$ is the distance projected on $\vunit{z}$ between a point and the center of the rivulet (the neutral axis of the beam),
and $F$ is the force, orthogonal to the cross-section, that compresses or elongates a non-neutral axis of the beam.
Indeed, under the bending deformation,
a non-neutral axis of the beam of initial length $\dd{x}$ and situated at a distance $\delta{z}$ of the center of the rivulet,
is subject to a relative deformation $\frac{\dd{(\dd{x})}}{\dd{x}} = -\frac{\delta z}{R_z}$, as is shown in figure~\ref{fig:drawing_bending}~(center).
The curvature radius $R_z = a/\kappa_z$ is defined as the inverse of the curvature of the path $\kappa_z\approx \partial_{xx}z$.
Hence, in order to close the problem,
we only need an equation relating the force $\dd{F}$ to the deformation of the beam $\frac{\dd{(\dd{x})}}{\dd{x}}$.

In the case of solid, elastic beams, this relationship is given by Hooke's law:
\begin{align}
    \sigma_{ii} = E\,\varepsilon_{ii}\ ,
\end{align}
where the $\sigma_{ij}$ are the components of the stress tensor,
and $\varepsilon_{ij}$ the ones of the strain (deformations) tensor.
This leads to the expression
\begin{align}
    \dd{F} &= E\, \frac{\dd{(\dd{x})}}{\dd{x}}\, \dd{\area}
    = - E\,\delta_z\, \kappa_z\, \dd{\area}
\end{align}
where $E$ is the Yong modulus of the material.
The bending moment is then proportional to the curvature of the beam $\kappa_z$.

In our case, since the rivulet is liquid, Hooke's law obviously can't apply.
However, it is possible to establish a formal analogy between the effect of elasticity in solid media and
the consequences of viscosity in liquids.
For a newtonian fluid, one can write
\begin{align}
    \sigma_{ii} = 2\,\rho\,\nu\,\dv{}{t}\varepsilon_{ii}\ ,
\end{align}
where $\dv{t} = \partial_t + \vb{u}\bcdot\bnabla$ corresponds to the total derivative,
that combines a local contribution and an advective contribution.
It is thus legitimate to ``replace'' $E$ by $2\,\rho\,\nu\,\dv{}{t}$:
using this analogy allows one to explain the buckling of viscous columns under compression~\cite{lemerrer2012},
the folding of liquid sheets~\cite{suleiman1981, cruickshank1981, howell1996, ribe2003},
the coiling of liquid filaments~\cite{mahadevan1998, ribe2012},
or the pattern generated by thin fluid threads falling onto a moving belt~\cite{brun2015}.

The correspondence was first noticed by Stokes~\cite{stokes1845}:
when deriving the equations for fluid mechanics that were previously obtained by Poisson,
he made the remark that it was possible to retrieve the equations for solid mechanics by making this replacement.
The same remark was also made later by Rayleigh~\cite{rayleigh1877}.

The most surprising, and in some sense counter-intuitive, feature of this analogy is that
instead of having the bending moment directly proportional to the curvature,
it is in the case of liquid beam proportional to the \textsl{rate of change} of the curvature~\cite{buckmaster1973, buckmaster1975, ribe2006, lemerrer2012}.

Following this approach, we obtain for our case $\dd{F} = -2\,\rho\,\nu \,\dv{\kappa_z}{t}\, \delta z\, \dd{\area}$,
and so
\begin{align}
    C = 2\,\rho\,\nu\,\dv{\kappa_z}{t}\iint_{(\area)}(\delta z)^2\dd{\area}
    = 2\,I_{y}\,\rho\,\nu\,\dv{\kappa_z}{t} \qqtext{, where} I_{y}\defeq \iint_{(\area)}(\delta z)^2\dd{\area}
\end{align}
is the {planar quadratic moment of area} of a transverse section of the rivulet.

Hence, the rivulet is subject to a bending force per unit length
\begin{align}
    \vb{f}_b  =  -\partial_{xx} C\,\vunit{n} = -\rho\,\area\,\mu\,B\,\partial_{xx}(\partial_t + \vb{u}\bcdot\bnabla){\kappa_z}\,\vunit{n}
    \qqtext{, where} B \defeq 24\,\frac{b^2}{\area}\frac{\mu_\infty}{\mu}\,I_{y}
\end{align}
 is the (viscous) bending modulus of the beam.
The analytical computation of the value of $I_{y}$ is presented in annex~\ref{sec:quadmom}.

\subsection{Final model}

Finally, we obtain the full dynamical equation
\begin{align}
    \label{eq:nsvectoriel}
    (\partial_t + \vb{u}\bcdot \bnabla) \vb{u}
    =&\ \vb{g} - \mu\,\vb{u}+ \qty(\vcap^2\,\kappa - \mucl\,\vb{u}\bcdot\vunit{n} - \mu\, B\, \partial_{xx}(\partial_t + \vb{u}\bcdot\bnabla)\kappa)\,\vunit{n}
\end{align}
as well as the kinematic condition
\begin{align}
    \label{eq:condlimit}
    \dv{z}{t} = (\partial_t + \vb{u}\bcdot\bnabla)z = \vb{u}\bcdot \vunit{z}
\end{align}
and the system is closed by the mass conservation equation
\begin{align}
    \label{eq:masscons}
    \partial_t \area = \bnabla \bcdot (\area\,\vb{u}) \qqtext{or} (\partial_t + \vb{u}\bcdot\bnabla)\area = -\area \, \bnabla \bcdot \vb{u}\ .
\end{align}

\section{Linear instability mechanism} \label{sec:linear}

Now that we have a mathematical model for the rivulet behavior,
let us use it and develop and understanding of the meandering instability.

\subsection{Linearized equations}

The base state (``order 0'') of the system consists of $z = 0$ and $\vb{u} = u_0\,\vunit{x}$ with $u_0 \defeq g / \mu$.
Since we made the hypothesis of small deformations we can now write, at first order, the linearized dynamical equation:
\begin{align}
(\partial_t + u_0\,\partial_x)^2 z
&= \vcap^2\,\partial_{xx} z - \mu\,(\partial_t + u_0\,\partial_x)(1 + B \partial_{x}^4) z
- {\mucl}\,\partial_t z\ .
\end{align}
It is interesting to place ourselves in the reference frame in which the fluid is at rest,
that is advected downward at $u_0$,
since it is the natural reference frame for the capillary waves.
We obtain a noticeably simpler expression
\begin{align}
    \label{eq:disprel_noadim_advected}
\partial_{tt}z - \vcap^2\,\partial_{xx} z
&= - \qty(\mu + \mucl + \mu \,B\,\partial_x^4)\,\partial_t z
+ u_0 {\mucl}\,\partial_x z
\end{align}
The viscosity ratio $\mu/\mucl$, that compares damping in the bulk to the supplementary dissipation due to the moving menisci,
is always inferior to one.

We make the equation dimensionless, using $\vcap$ as velocity scale and $b$ as length scale, forcing $\vcap/b$ as the inverse-time scale.
The bending modulus and friction coefficients are rescaled appropriately: $\Badim \defeq B / b^4$, $\muadim \defeq \mu\,b/\vcap$ and $\mucladim \defeq \mucl\, b/\vcap$.
The only other parameter is now the capillary Mach number $\mach \defeq u_0/\vcap =\sqrt{\text{We}}$,
which corresponds to the square root of the typical Weber number in this problem.
\begin{align}
    \label{eq:dimensionlesslinear}
    \partial_{tt}z - \partial_{xx} z
    &= - \qty(\muadim + \mucladim + \muadim \,\Badim\,\partial_x^4)\,\partial_t z
    + \mach\,\mucladim\,\partial_x z
\end{align}
Let us first try to qualitatively understand equation~\eqref{eq:dimensionlesslinear}.
On the left-hand side, we have a wave equation, indicating that the solutions will take the form of propagating and counter-propagating transverse capillary waves.
On the right-hand side, we have first a dissipative term, that damps indiscriminately both kind waves.
The fourth order space derivative indicates that the damping is wavelength-dependent : higher wavenumbers are more strongly attenuated.
Last, the rightmost term is a spatially anti self-adjoint term : it introduces an imaginary component in the dispersion relation.
This breaks the symmetry between propagating and counter-propagating waves :
it adds supplementary damping for propagating waves (for which $\partial_x \propto -\partial_t$),
but it functions as an amplification term for counter-propagating waves (for which $\partial_x \propto \partial_t$).
If this last term becomes greater than the damping (in norm),
it can lead to the instability of one of the two type of waves,
and the over-attenuation of the other.

\subsection{Instability criterion}

In order to obtain rigorously the instability criterion, we can use a Fourier-Laplace décomposition by using the ansatz
$z = \underline{z}\, e^{s\, t-i\, k\, x}$, with $k$ real and $s$ complex.
The waves are then unstable if there existe a real $k$ for which the temporal growth rate $\Re(s)$ is positive.
We then obtain the dispersion relation
\begin{align}
    s^2 + \mutotadim s + k^2 + i\,  \mach\,\mucladim\,k
    &= 0 \qqtext{with} \mutotadim = \muadim + \mucladim + \muadim \,\Badim\,\partial_x^4
\end{align}
which admits the solutions
\begin{align}
        s^{\pm} = \frac{1}2 \qty(-\mutotadim \pm \delta) \qqtext{with}
    \delta^2 = \mutotadim^2- 4\,k^2 - 4\,i\,k\,\mach\,\mucladim \qqtext{and} \Re(\delta) > 0 \ .
\end{align}
It is obvious that the mode $s^-$ is over-damped and we shall no longer be interested in it.
On the contrary, it is possible for the $s^+$ mode to display a positive real part,
meaning it is exponentially amplified.
The instability criterion is then $\abs{\Re(\delta)} > \mutotadim$.
The main difficulty now is to transform this criterion into a relationship between the parameters.
This can be done at the expense of tedious algebra tricks, which it is difficult to attribute any meaningful signification to.
Finally, one obtains the simple instability criterion:
\begin{align}
    \label{eq:instabilitycriterion}
    \mach > \frac{\mutot}{\mucl}
    \qqtext{i.e.}
     \frac{u_0}{\vcap} > 1 + \frac{\mu}{\mucl} + \frac{\mu}{\mucl}\,\Badim\,k^4
    \ .
\end{align}

The rivulet is thus unstable if the speed at which the rivulet falls down $u_0$ becomes greater than the celerity of transverse capillary waves $\vcap$.
The critical ratio for this instability depends on the supplementary friction brought by the menisci displacement.
In particular, when $\mucl \gg \mu$, transverse rivulet movement are seriously penalized,
the contact line is ``almost pinned'' (truly pinned contact lines imply pinning forces, which have not been considered here),
and the instability arises as soon as $u_0$ and $\vcap$ match.
A more in-depth physical interpretation of this criterion is done in section~\ref{subsec:phsicalinterp}.

Thanks to the incorporation of viscous bending into the model,
the threshold value for $\mach$ now increases with the fourth power of the wavenumber $k$.
The rivulet is then stable with respect to small wavelength perturbations,
in contrast with the findings of previous studies~\cite{daerr2011}.
Without this critical ingredient, the instability would trigger for all wavenumbers at the same time.

One could of course argue that there exists instabilities for which all wavenumbers are unstable.
Such is the case of the jet pinching (Rayleigh-Plateau) instability,
or the Kelvin-Helmholtz instability in the absence of surface tension.
However, the problem here is even deeper:
at high wavenumbers ($k\gg 1$), in the absence of bending, one can write
\begin{align}
    \delta^2 \sim -4\, k^2- 4\,i\,\mach\,\mucladim\, k \approx \qty( M\,\mucladim - 2\,i\,k)^2
    \qqtext{, hence} s^+ \approx \frac{\mucladim}{2}\qty(M - \frac{\mu + \mucl}{\mucl}) - i\,k\ .
\end{align}
This means that in the absence of viscous bending, when $\mach > 1 + \frac{\mu}{\mucl}$,
the real part of $s^+$ not only is positive but \textsl{does not depend on $k$} at high wavenumbers.
All wavenumbers would then be amplified indiscriminately: this is a completely unphysical picture.

\subsection{Approximate growth rate}

\begin{figure}[h]
    \centerline{\includegraphics[scale=1]{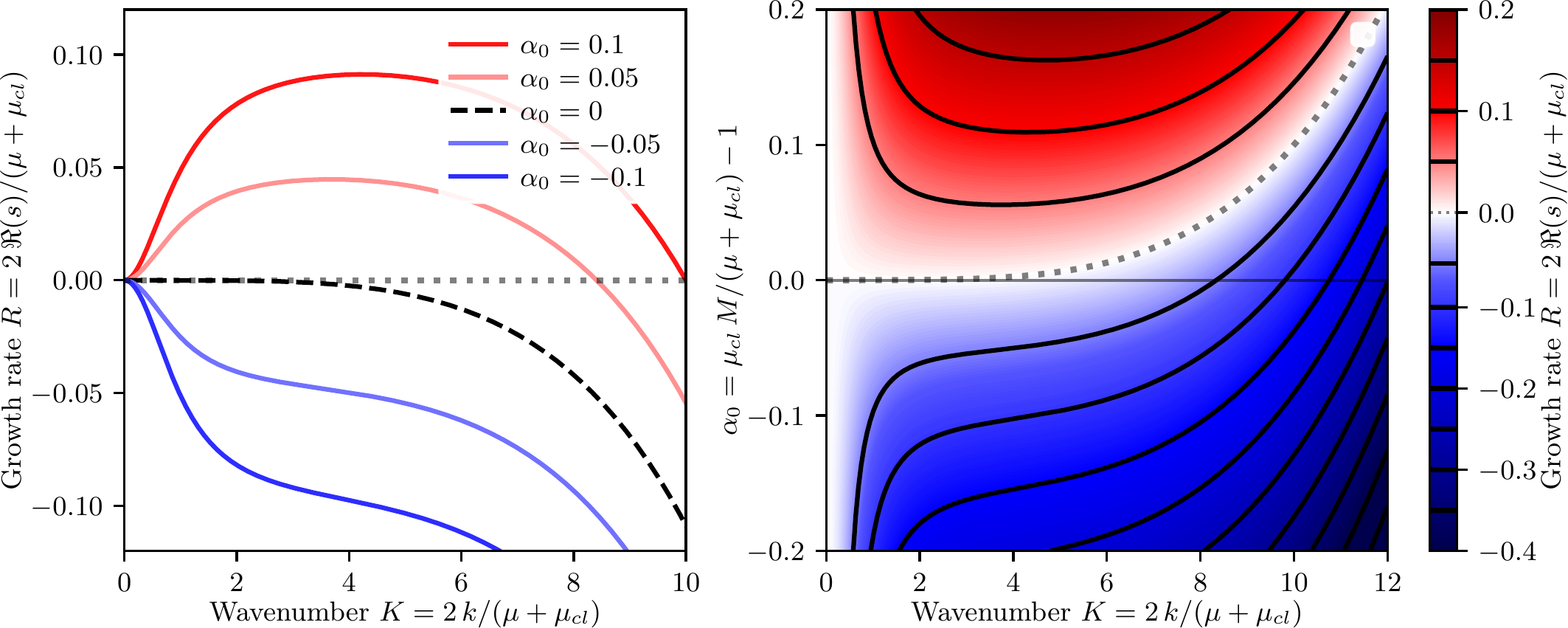}}
    \caption{
        (left) Dimensionless growth rate (rescaled) as a function of the (rescaled) wavenumber,
        for varying values of the instability parameter $\crit_0$.
        \\
        (right) In color scale, the dimensionless growth rate as a function of the wavenumber $K$ (horizontal axis)
        and the instability parameter $\crit_0$ (vertical axis).
        Both plots are with $\mathcal{B} = 10^{-5}$.
    }
    \label{fig:SM_disprel_plot_exact}
\end{figure}

The exact growth rate for the unstable mode is
\begin{align}
    \label{eq:growthrate}
    s^+ = \frac{\mutotadim}2 \qty(-1 + \sqrt{\qty(1- i\,\frac{2\,k}{\mutotadim})^2 - 2\,i\,\crit\,\frac{2\,k}{\mutotadim}})
\end{align}
where we introduced the new variable
\begin{align}
    \crit = \mach\,\frac{\mucl}{\mutot}-1 = \frac{\mucl\,\mach}{\mu + \mucl + \mu \Badim k^4} -1
\end{align}
to represent the instability criterion: when $\crit < 0$, the straight rivulet is linearly stable, and when $\crit >0$ the instability develops.
Plots of this growth rate can be found in figure~\ref{fig:SM_disprel_plot_exact}.
The expression of the growth rate~\eqref{eq:growthrate} is quite complex,
and makes it uneasy to interpret the growth (especially since $\mutotadim$, and thus $\crit$ varies with $k$).

In this section, we show that it is possible to obtain a simpler expression for he growth rate, 
which is far easier to interpret, using two key assumptions:
\begin{enumerate}[(i)]
    \item We are near the instability threshold, i.e. $\abs{\crit} \ll 1$;
    \item The behavior of the rivulet is not dominated by viscous bending, which stays marginal, i.e. $\mu \Badim k^4 \ll \mu + \mucl$.
\end{enumerate}
At the threshold exactly, when $\crit = 0$, we have $s^+ = -i\,k$: the waves are neither amplified nor damped,
and travel upward at the capillary speed (which is $1$ is our dimensionless units).
Near the threshold, we can thus write $s^+  = r -i\,k (1 + \cphi)$, where $r \ll 1$ is the growth rate,
and $\cphi \ll 1$ is the supplementary phase speed of the perturbation.
Both quantities are real numbers, and assumption (i) guarantees that $r \ll 1$ and $\cphi \ll 1$.

Moreover, assumption (ii) translates to $\mutot \approx \mu + \mucl$, and
\begin{align}
    \crit \approx \crit_0 - \frac{\mu}{\mu+\mucl}\, \Badim\, k^4 \qqtext{with} \crit_0 = \frac{\mach}{1+\mu/\mucl}-1 \ .
\end{align}

\begin{figure}[h]
    \centerline{\includegraphics[scale=1]{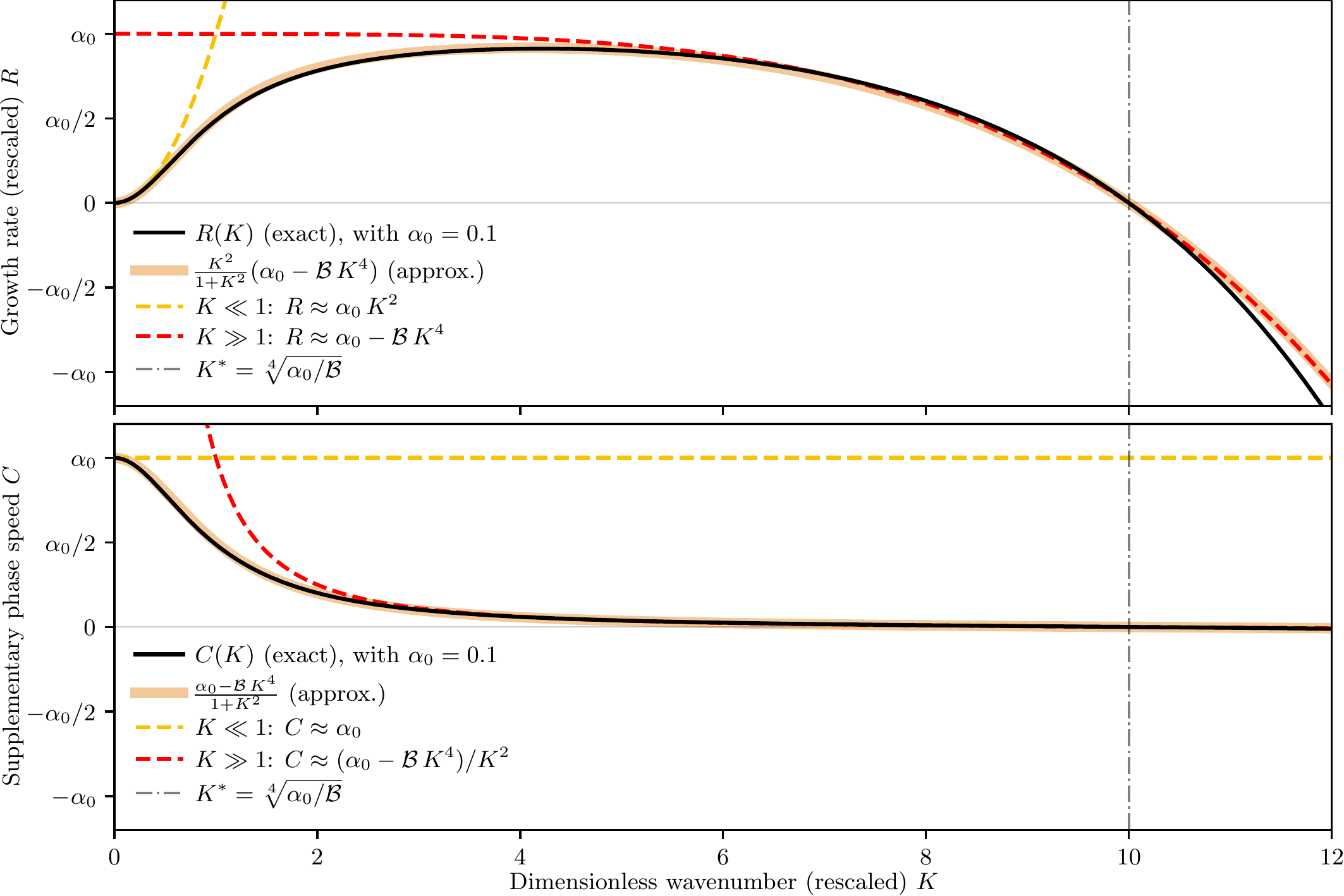}}
    \caption{
        Growth rate $R$ (top) and supplementary phase speed $c$ (bottom) as a function of the wavenumber $K$, for $\crit_0 = 0.1$ and $\mathcal{B} = 10^{-5}$.
        The yellow and red dashed lines represent the low-wavenumber and high-wavenumber regimes, respectively.
        The thick orange line represent the approximate solution given in~\eqref{eq:approxdisprel},
        to be compared with the black line which is the exact solution computed from~\eqref{eq:growthrate}.
    }
    \label{fig:SM_disprel_plot_approx}
\end{figure}

Injecting this ansatz into the dispersion relation,
and neglecting terms of second order in $r$, $\cphi$ and $\mu\,\Badim\, k^4/(\mu+\mucl)$,
we obtain two equations : one on the real and one on the imaginary components,
which reduce to
\begin{align}
    \label{eq:approxdisprel_beforerescale}
    r & = \frac{2\, k^2\, \cphi}{\muadim+\mucladim}
    \qqtext{and}
    \cphi = \crit_0 - \frac{\mu}{\mu+\mucl}\, \Badim\, k^4 - \frac{2\,r}{\muadim+\mucladim}
    \ .
\end{align}
We choose to use rescaled variables $R = 2\,r/(\muadim+\mucladim)$, $K=2\,k/(\muadim+\mucladim)$, and $\mathcal{B} = \muadim\,(\muadim+\mucladim)^3\,\Badim/16$.
These allows for simpler expressions, which better highlights the physics of the system.
Combining equations~\eqref{eq:approxdisprel_beforerescale},
one then obtains
\begin{align}
    \label{eq:approxdisprel}
    R &= \frac{K^2}{1 + K^2}\qty(\crit_0 - \mathcal{B}\, K^4)
    \qqtext{and}
    \cphi = \frac{\crit_0 - \mathcal{B}\, K^4}{1+ K^2}
\end{align}
which are plotted in figure~\ref{fig:SM_disprel_plot_approx}.

The absence of odd powers of $K$ can be linked simply to the fact that the problem is symmetric relative to the sign of $\partial_x z$.

For very long wavelengths, $K \ll 1$ and $\mathcal{B}\,K^4\ll \crit_0$, hence $c = \crit_0$ and $R = \crit_0\, K^2$.
The growth rate thus evolves quadratically with the wavenumber.
Note that the fact that $R(K\rightarrow 0) = 0$ is imposed by the fact that the system is invariant under the transformation $z \rightarrow z_0$.
The perturbations are slightly faster than the capillary velocity in the advected frame of reference.

The growths rate changes sign when $\mathcal{B} K^4 \sim \crit_0$.
This can correspond to two different asymptotic regimes:
\begin{itemize}
    \item In the bending-dominated regime, which happens extremely close to the instability threshold when $\crit_0 \ll \mathcal{B}$,
    the change of sign can happen when $K \ll 1$.
        Hence, $R \approx \crit_0 \, K^2 - \mathcal{B}\, K^6$: viscous bending acts as an extremely strong sixth-order dissipation term,
    which drastically cuts small-wavenumber perturbations.
    In practice, this regime is never observed, and we will no longer take it into account in the following.
    Indeed, for experimentally sound values of the parameters, one finds $\mathcal{B}$ to be of order $10^{-5}$,
    and it is not experimentally feasible to be this close to the instability threshold.
    Hence this regime is deemed not relevant experimentally.

    \item When $\crit_0 \gtrsim \mathcal{B}$, the crossover happens while $K \gg 1$.
    After the initial quadratic growth, $R$ stabilises around the value $\crit_0$, and $c$ falls quickly to 0.
    Then, when $\crit_0 \approx \mathcal{B}\,K^4$, the growth rate falls and become negative
\end{itemize}
In the second scenario, the main effect is that the growth rate evolution $R(k)$ reaches a stable plateau at $\crit_0$.
This explains why the instability shows low linear selectivity and accepts such a huge bandwidth.

\subsection{Convective nature of the instability}

We can now try to determine if the meandering instability is absolute or convective in nature.
This can be done by searching for saddle points the complex $k$ plane~\cite{huerre1990, duprat2007},
and applying the Briggs–Bers criterion~\cite{briggs1964, bers1975}.
To do so, we place ourselves back into the laboratory frame of reference,
and we use of the normal mode ansatz $z = \underline{z} e^{i(\omega\, t-k\, x)}$,
where now both $\omega$ and $k$ can be complex.
We thus obtain the dispersion relation
\begin{align}
    D(\omega, k) = (\omega - \mach\,k)^2 - k^2 - i \muadim (\omega- \mach\,k)\,(1 + \Badim k^4) - i\,\omega\,\mucladim
    &= 0 \ .
\end{align}

In order to obtain analytical results,
we only consider the case where the small wavelength dissipation can be neglected, translating as $\Badim\,k^4 \ll 1$.
Taking this limit allows us to do the computations analytically,
and is reasonable if the end result is that the instability is purely convective:
since the $\Badim\,k^4$ term only brings supplementary damping,
we do not expect that by itself it is able to change the status of the instability from convective to absolute.

We now search for a saddle points $(\omega_0, k_0)$ in the complex $k$ plane, defined by the relation
\begin{align}
    \dv{k_0}{\omega_0} = 0 \Rightarrow \eval{\pdv{D}{k}}_{\omega_0,k_0} &= 2\,\mach\,(\mach\,k_0-\omega_0) -2 k_0 + i\,\mach\,\muadim= 0
    \\
    \qqtext{which leads to} k_0&= \frac{\mach}{\mach^2 - 1} \qty(\omega_0 - i \frac{\muadim}2)
\end{align}
and we want to determine the sign on the imaginary part of $\omega_0$,
which is defined by $D(\omega_0,k_0) = 0$, i.e.
\begin{align}
{\omega_0}^2 - i\,\omega_0\,\qty(\muadim + \mucladim \qty(1-\mach^2))  - \qty(\frac{\mach\,\mucladim}{2})^2 &= 0 \ .
\end{align}
We then obtain that
\begin{align}
    \omega_0
    &= \frac{\muadim + \mucladim \qty(1-\mach^2)}{2}\qty(i \pm \sqrt{\qty(\frac{\mach\,\mucladim}{\muadim + \mucladim \qty(1-\mach^2)})^2 - 1})
\end{align}
which always has positive imaginary part.
With our sign convention, it means that perturbations wih zero group velocity are systematically attenuated.
Hence, the instability is purely convective and there is no regime of parameters for which it is absolute in the laboratory frame of reference.
This also justifies our simplifying assumption $\Badim\,k^4\approx0$.

With similar line of reasoning, one can also show that the instability as also always convective in the
frame of reference advected with the fluid a speed $u_0$ (in which perturbations move upwards at speed $\vcap$).
Finally, it is also possible to show that in the frame of reference moving with the perturbation at speed $u_0 - \vcap$ relative to the lab,
the instability is absolute provided that $M > 1 + \mu/\mucl$,
which unsurprisingly corresponds to the instability criterion~\eqref{eq:instabilitycriterion}.

\section{Physical interpretation}\label{sec:physics}

In this section we are interested in the physical interpretation of the instability: what physical effects are causing it?
In our system, inertia, capillarity, viscosity and meniscus friction all play a role, so that it can seem difficult to disentangle them
and establish the effect that is mainly responsible for the growth of a perturbation above the threshold.
To simplify the following discussion, in this section we chose to not take bending into account,
as it only plays a role in limiting the growth bandwidth but does not influence the instability criterion.
This very criterion can be written
\begin{align}
    \mucl \, (u_0 - \vcap) > \mu \, \vcap \qqtext{or} \frac{u_0/\vcap}{1 + \mu/\mucl} > 1 \qqtext{or simply} \alpha_0 > 0 \ .
\end{align}

\subsection{Inertial or viscous instability?}

In 2011, Daerr et al.~\cite{daerr2011} interpreted this criterion as a competition between centrifugal forces and surface tension.
They argued that in the referential frame of the waves, moving at celerity $u_\phi$,
the fluid is moving at $u_0 - u_\phi$, and that the rivulet is unstable
when in this referential centrifugal force $\rho (u_0 - u_\phi)^2$ overcomes capillary restoring force $\rho \vcap^2$.
The instability would in this case be driven by a change of sign of the restoring force,
which would become negative, much like in the case of the Kelvin-Helmholtz inertial instability.

We think that this line of reasoning at worst inadequate, and at best insufficient to explain the instability.
In the previous section, we showed that the capillary waves propagate at a celerity $(1 + c) \vcap$ in the frame of reference advected at $u_0$,
meaning that the phase speed of the perturbations in the lab frame of reference is $u_\phi= u_0 - (1 + c) \vcap$,
and that the speed of the fluid particules in the perturbation frame of reference is $\Delta u = u_0 - u\phi = (1 + c) \vcap$.
In that regard, if the instability was inertial in nature,
the growth rate should increase (or at the very least correlate) with $\abs{c}$.
However this is not the case at low wavenumbers: $c$ is maximum at $k=0$, where the growth rate is null,
and and $k$ augments, $c$ decays and the growth rate augments.
The argument for an inertial instability fails even worse at high wavenumbers,
where $c$ decays to zero, while the growth rate stays constant (in the absence of bending).

This can be even better highlighted by looking at the evolution of an ansatz where $c=0$.
For this, we consider a low-viscosity situation,
where $\mu = \eps\, \mu^*$ and $\mucl = \eps\,\mucl^*$ with $\eps \ll 1$,
and we consider the simplified solution
\begin{align}
    z(x, t) = A(T) f(x + \vcap\, t) \qqtext{where} T = \eps t \qqtext{is a slow time scale}
\end{align}
and where $f$ is an unknown function representing the counter-propagating wave in the advected frame of reference.
The parameter-free form of the  criterion derived by~\cite{daerr2011} can be written $(\Delta u)^2 > (\vcap)^2$ i.e. $\abs{c} > 0$:
here we have $c=0$, so there should be no growth of the perturbation.
However, under our assumptions, the dispersion relation in the advected frame of reference~\eqref{eq:disprel_noadim_advected} becomes
\begin{align}
    \partial_{tt}z + \eps\,\partial_{tT}z - \vcap^2\,\partial_{xx} z
    &= - \eps\,\qty(\mu^* + \mucl^*)\,\partial_t z
    + \eps\,u_0 {\mucl^*}\,\partial_x z + O(\eps^2)
    \\
    \vcap\,f'\, \partial_T A
    &= - \qty(\mu^* + \mucl^*)\, A\, \vcap\,f'
    + u_0 {\mucl^*}\, A\,f' \qqtext{at order $\eps$}
    \\
    \label{eq:nonlinearansatz}
    \partial_T A &= \qty[(u_0 - \vcap){\mucl^*} - \mu^* \vcap] A
\end{align}
Hence, a solution which is not slowed down by friction, for which capillary waves travel at $\vcap$,
is linearly amplified despite being subject to no supplementary inertial force in the reference frame of the perturbation!

In equation~\eqref{eq:nonlinearansatz}, the stabilizing effect is proportional to the bulk viscosity coefficient $\mu$,
while the main driving component for the instability seems to be $(u_0 - \vcap)\mucl$,
which correspond to viscous meniscus friction (since the meniscus moves a speed $u_0 - \vcap$ relative to the lab).
This leads us to believe that the main force responsible for the destabilization is indeed meniscus friction,
and that the spontaneous meandering instability is driven by the change of sign of viscosity,
(rather than elastic restoring force in the inertial case).

\subsection{Physical interpretation}\label{subsec:phsicalinterp}

\begin{figure}[h]
    \centerline{\includegraphics[scale=1]{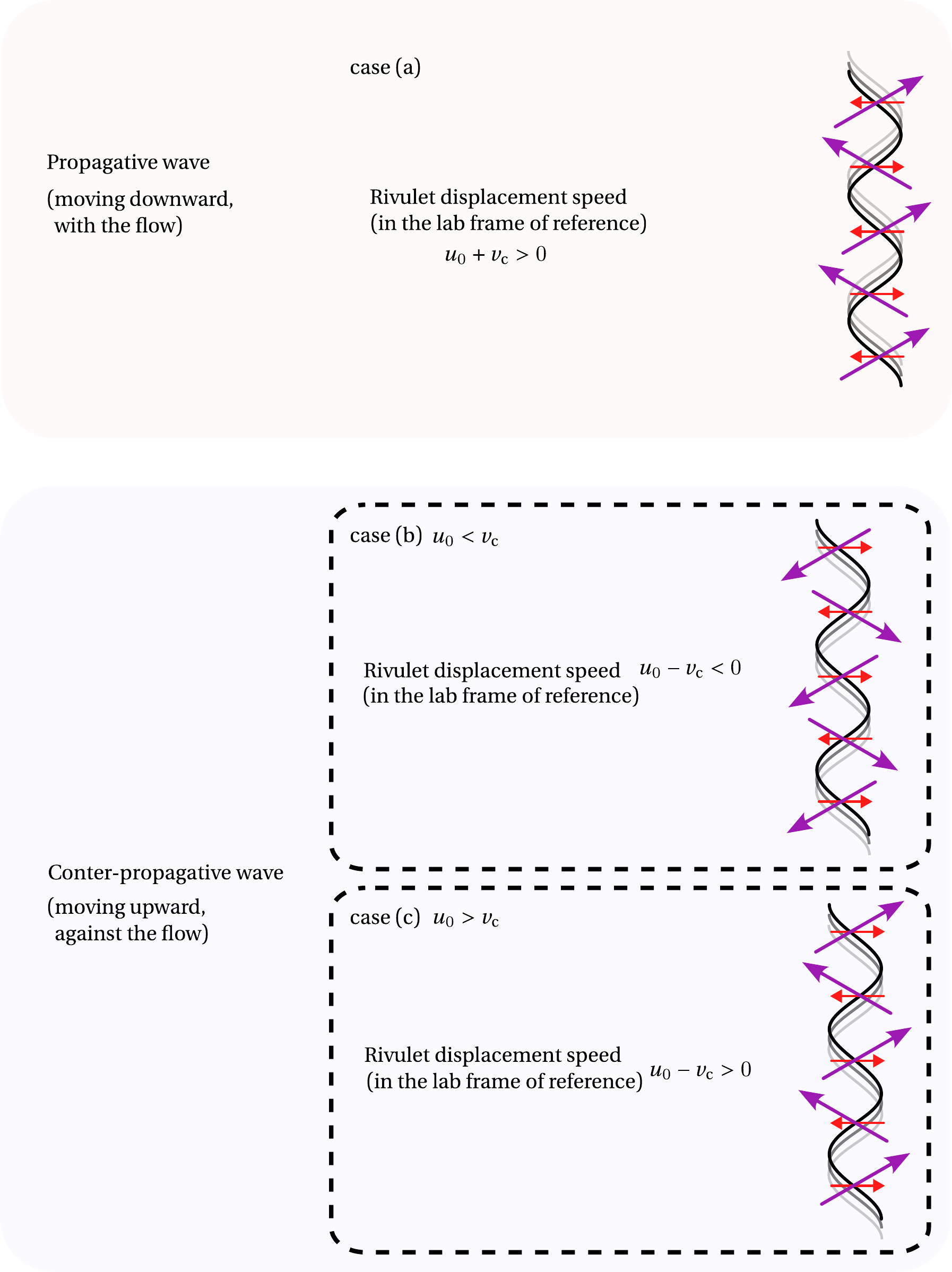}}
    \caption{
        Physical illustration of the forces at play.
    }
    \label{fig:schema_meca_instab}
\end{figure}

How can meniscus friction, which is at first sight a resisting force,
which opposes the rivulet movement, be responsible for the growth of a perturbation?
In this section, we provide a qualitative explanation of this surprising mechanism,
in order to get a physical idea of the role of each of the forces,
using the illustrative drawings on figure~\ref{fig:schema_meca_instab}.
To simplify the physical discussion, we neglect here the effect of bending,
as well as the effect of the bulk viscosity $\mu$.

On this figure,
all drawings on the right are represented in the free fall frame of reference,
that is advected with the fluid at $u_0$,
in which near the threshold the fluid particules do not move up or down.
Since capillary waves propagate on the rivulet at celerity $\vcap$,
the fluid particules move aither to the left or to the right, in order to propagate the deformation.
The instantaneous speed of the fluid particles is shown by red arrows.
The past state of the rivulet is shown with varying levels of gray.

The meniscus friction force is represented by purple arrows:
it is normal to the rivulet path, by definition,
and proportional to the rivulet displacement speed \textsl{in the laboratory frame of reference}.
When this force is in the opposite direction as the speed of the fluid particles,
it works as an attenuation force, diminishing the amplitude of the waves.
When in the contrary the red and purple arrows have a positive scalar product,
friction amplifies the movement.

In case (a) we consider a propagating wave, that is moving downwards in the advected frame of reference.
In the laboratory frame of reference, its speed is $u_0 + \vcap > 0$: it is also moving downwards,
the waves go in the same direction as the flow.
This downward movement generates an upward friction force, which fixes the direction of the meniscus friction:
this force opposes the instantaneous speed of the fluid particles,
meaning that the friction is purely resistive and the wave is strongly attenuated.

In case (b) we consider a counter-propagating wave, that is moving upwards in the advected frame of reference,
with $u_0 < \vcap$.
In the laboratory frame of reference, its speed is $u_0 - \vcap < 0$: it is also moving upwards,
the waves go against the flow and the propagation speed is greater than the free-fall speed,
so that the waves move upstream.
This upward movement generates a downward friction force, which fixes the direction of the meniscus friction:
this force opposes the instantaneous speed of the fluid particles,
meaning that the friction is again purely resistive and the wave is attenuated.

In case (c) we consider a counter-propagating wave, that is moving upwards in the advected frame of reference,
this time with $u_0 > \vcap$.
In the laboratory frame of reference, its speed is $u_0 - \vcap > 0$: it is this time moving downward,
the waves go against the flow but the propagation speed is smaller than the free-fall speed,
so that the waves move downstream in the laboratory frame of reference.
This downward movement generates an upward friction force, which fixes the direction of the meniscus friction:
this force is in the same direction as the instantaneous speed of the fluid particles,
meaning that this time friction works with the wave propagation,
this triggers the instability and the wave is amplified.

It is thus possible to understand the physical mechanism of amplification by contact-line friction:
this friction force always opposes the movement of the meniscus in the reference frame of the laboratory,
but under the condition $u_0 > \vcap$,
it can accompany and amplify capillary waves in the advected frame of reference.
Indeed, when $\mu = 0$, the instability criterion reduces to $M = u_0 / \vcap > 1$.

\section{Multiple scales expansion}\label{sec:multiscale}

In this paper, we worked in the linear approximation,
but the experimental spontaneous meandering exhibits strong nonlinear behavior~\cite{daerr2011}.
In order for future nonlinear studies to be able to use our work, and in particular to benefit from the simplification
of the expression of the growth rate at which we arrived, we show that we can reproduce the
approximated expression for the linear growth rate given in~\eqref{eq:approxdisprel} by using a multiple scales expansion.

We use the rescaled variables $\xi = \eps\,(x + (V+\eps\,v)\,t)$, where $V$ is the propagating speed of the wave and
$v$ is the small, supplementary drift speed of the pattern
(note the sign, which indicates that we consider a counter-propagating wave if $V > 0$) ;
and $\tau = \eps^2 t$ which is a slow time scale.
We also make the assumption that we are in a low-dissipation situation,
i.e.~we consider that $\mu$ and $\mucl$ are of order $\eps \ll 1$.
In order to keep bending into the model, we also make the assumption that $B$ scales with $\eps^{-3}$

We start from equation~\eqref{eq:dimensionlesslinear}
\begin{align}
    \partial_{tt}z - \partial_{xx} z
    &= - \qty(\muadim + \mucladim  + \muadim B \, \partial_{xxxx})\,\partial_t z
    +\mucladim\, \mach\, \partial_x z
\end{align}
At order $\eps^2$, we have
\begin{align}
\label{eq:multiscale_epsilon2}
(V^2 - 1)\,\partial_{\xi\xi}z
    &= \mucladim \qty[
     \mach - V  \qty(1 + \frac{\mu}{\mucl})] \partial_{\xi} z \ .
\end{align}
The system is invariant under the reflection symmetry $x, t \rightarrow -x, -t$,
which here corresponds to $\xi\rightarrow -\xi$.
This means that both sides of equation~\eqref{eq:multiscale_epsilon2} must be null.
For the left-hand side to be zero, we need $V = \pm 1$.
And the nullity of the right-hand side imposes $V = 1$.
We choose to write $\mach = \qty(1 + \frac{\mu}{\mucl}) (1+ \eps \crit_0)$,
making the supplementary assumption that we are near, but not exactly at,
the threshold.
Doing sends the right-hand side to the next order.

We can then look at order $\eps^3$
\begin{align}
    2V\partial_{\tau\xi}z + 2 V v \partial_{\xi\xi}z
    &= - \qty(\muadim + \mucladim )\,\qty[ \partial_\tau + (v - \crit_0) \partial_{\xi}] z
    -\muadim V B \partial_{\xi\xi\xi\xi\xi} z
\end{align}

By again taking into account the symmetry $\xi \rightarrow -\xi$,
we can separate the odd and even parts of the equation:
\begin{subequations}
    \begin{align}
        2V\partial_{\tau\xi}z
        &= -  \qty(\muadim + \mucladim )(v - \crit_0) \partial_{\xi} z -\muadim V B \partial_{\xi\xi\xi\xi\xi} z
        \\
         2 V v \partial_{\xi\xi}z
        &= - \qty(\muadim + \mucladim )\, \partial_\tau  z
    \end{align}
\end{subequations}
Finally, one obtains the master equation
\begin{align}
    \qty[ \qty(\frac{\muadim + \mucladim}{2})^2 - \partial_{\xi\xi}] \partial_{\tau} z &= - \crit_0\frac{\muadim + \mucladim}{2}\partial_{\xi\xi}z
    + \frac{\muadim}{2} B \partial_{\xi\xi\xi\xi\xi\xi} z
\end{align}

Which we can put in a normalized form:
\begin{align}
    \qty[ 1 - \partial_{X}^2] \partial_{T} z &= - \crit_0\, \partial_{X}^2 z + \mathcal{B}\, \partial_{X}^6 z
\end{align}
with $X = (\muadim + \mucladim)\,\xi/2$ and $T = (\muadim + \mucladim)\,\tau/2$.

Following the same line of thoughts, but using a different ansatz
(for example, $z = \eps \zeta(\xi, \tau)$, $\xi = x + (V + \eps^2 v) t$ and $\tau = \eps^{2} t$),
one can add complexity and nonlinear terms into the model.

\section{Conclusion and perspective}\label{sec:conclu}

In this work,
we have addressed the long-standing problem of wavelength selection in the spontaneous meandering of liquid rivulets.
By incorporating viscous bending into the depth-averaged Navier-Stokes equations,
we have provided a robust linear stability model that successfully resolves the unphysical short-wavelength divergence found in previous studies.
This addition introduces a natural small-wavelength cut-off,
establishing viscous bending as the primary physical mechanism governing the characteristic scale of the meandering pattern.

Our results offer a fundamental reinterpretation of the instability’s driving force.
While meandering was earlier attributed to inertial effects,
drawing parallels to centrifugal instabilities,
our analysis shows that menisci friction is instead the primarily driver of the instability.
Furthermore, through the application of the Briggs–Bers criterion,
we have shown that this instability is purely convective in the laboratory frame,
confirming the experimental observation that the meandering patterns are sensitive to upstream noise rather than representing a global breakdown of the flow state.

Finally, the derivation of the amplitude equations via multiple scales expansion provides a versatile mathematical framework for future research.
This opens a clear path toward investigating the highly nonlinear regime of meandering.

\paragraph{Acknowledgements} The authors would like to thank Laurent Limat for the deep physical insights they provided.

\paragraph{Data availability} The data used for this study are available upon reasonable request from the authors.

\paragraph{Open access} For the purpose of Open Access,
a CC-BY public copyright licence has been applied by the authors to the present document
and will be applied to all subsequent versions up to the Author Accepted Manuscript arising from this submission.

\clearpage

\section{ANNEX - Computation of $I_y$}\label{sec:quadmom}

We compute here the quadratic moment of a cross section of the rivulet,
with regard to the axis in the direction $\vunit{y}$ that passes through the rivulet center, where $\delta z = 0$.
The notations are the ones shown on fig.~\ref{fig:drawing_bending}~(right).

\begin{align}
    I_{y} &\defeq \iint_{(\area)} (\delta z)^2 \dd{\area}\\
    &= \int_{y=-b/2}^{b/2}\int_{\delta z=-z_m(y)}^{z_m(y)} (\delta z)^2 \dd{z}\dd{y} \qqtext{with} z_m(y) = \frac{w}2 + \frac{b}2 -\sqrt{\qty(\frac{b}{2})^2 - y^2}\\
    &= 4 \int_{y=0}^{b/2}\dd{y}\int_{\delta z=0}^{z_m(y)} (\delta z)^2 \dd{z}\\
    &= \frac{4}{3} \int_{y=0}^{b/2} \qty[z_m(y)]^3 \dd{y}\\
    &= \frac{4}{3} \qty(\frac{b}{2})^4 \int_{u=0}^{1} \qty[\frac{w}{b} + 1 - \sqrt{1 - u^2}]^3 \dd{u} \qqtext{using} y \eqdef \frac{b}{2}u\\
    &= \frac{4}{3} \qty(\frac{b}{2})^4 \int_{t=0}^{\pi/2} \qty[1 + \frac{w}{b} - \cos t]^3 \cos(t) \dd{t} \qqtext{using} u \eqdef \sin t \\
    &= \frac{b^4}{12} \qty[ \qty(1 + \frac{w}{b})^3 - \frac{3\pi}{4} \qty(1 + \frac{w}{b})^2 + 2 \qty(1 + \frac{w}{b}) - \frac{3\pi}{16} ]\\
    \qqtext{hence} B &\defeq 24\,\frac{b^2}{\area}\frac{\mu_\infty}{\mu}\,I_{y} = 2\,\frac{b^6}{\area}\frac{\mu_\infty}{\mu} \qty[ \qty(1 + \frac{w}{b})^3 - \frac{3\pi}{4} \qty(1 + \frac{w}{b})^2 + 2 \qty(1 + \frac{w}{b}) - \frac{3\pi}{16} ]
\end{align}

\end{document}